\documentclass[pra,aps,showpacs,twocolumn]{revtex4-1}
\usepackage{amsmath}
\usepackage{amssymb}
\usepackage{graphicx}
\usepackage{epstopdf}
\usepackage{easybmat}
\usepackage{color,soul}
\usepackage{times,txfonts}
\usepackage{ulem}
\usepackage{mathtools}
\newcommand{\ket}[1]{|#1\rangle}
\newcommand{\bra}[1]{\langle #1|}
\usepackage[bookmarks=false]{hyperref}
\hypersetup{colorlinks=true,citecolor=blue,linkcolor=blue,urlcolor=blue,pdfstartview=FitH,bookmarksopen=true}


\begin{document}

\title {Nonadiabatic holonomic quantum computation based on commutation relation }
\author{P. Z. Zhao}
\email{pzzhao@sdu.edu.cn}
\affiliation{Department of Physics, Shandong University, Jinan 250100, China}
\author{D. M. Tong}
\email{tdm@sdu.edu.cn}
\affiliation{Department of Physics, Shandong University, Jinan 250100, China}
\date{\today}

\begin{abstract}
Nonadiabatic holonomic quantum computation has received increasing attention due to the merits of both robustness against control errors and high-speed implementation. A crucial step in realizing nonadiabatic holonomic quantum computation is to remove the dynamical phase from the total phase. For this reason, previous schemes of nonadiabatic holonomic quantum computation have to resort to the parallel transport condition, i.e., requiring the instantaneous dynamical phase to be always zero. In this paper, we put forward a strategy to design nonadiabatic holonomic quantum computation, which is based on a commutation relation rather than the parallel transport condition. Instead of requiring the instantaneous dynamical phase to be always zero, the dynamical part of the total phase is separated from the geometric part and then removed by properly choosing evolution parameters. This strategy enhances the flexibility to realize nonadiabatic holonomic quantum computation as the commutation relation is more relaxed than the parallel transport condition. It provides more options for realizing nonadiabatic holonomic quantum computation and hence allows us to optimize realizations such as the evolution time and evolution paths.
\end{abstract}

\maketitle

\section{Introduction}

Practical applications of circuit-based quantum computation need to realize a universal set of accurately controllable quantum gates. However, the errors resulting from the imperfect control of a quantum system and the decoherence caused by the interaction between the quantum system and its environment inevitably affect quantum gates, which is the main obstacle to quantum computation. This practical issue motivates researchers to design quantum gates by utilizing the features of geometric phases \cite{Berry,Wilczek,Aharonov,Anandan,Jones,Zanardi,Duan,Wang,Zhu2002,Sjoqvist2012,Xu}.

Quantum computation based on nonadiabatic non-Abelian geometric phases \cite{Anandan} is known as nonadiabatic holonomic quantum computation \cite{Sjoqvist2012,Xu}. Since nonadiabatic non-Abelian geometric phases are only dependent on evolution paths but independent of evolution details, nonadiabatic holonomic gates possess a completely geometric property, being robust against control errors. Furthermore, nonadiabatic non-Abelian geometric phases avoid the long-run time required for adiabatic geometric phases and therefore nonadiabatic holonomic gates allow for high-speed implementation. Due to the merits of both robustness against control errors and high-speed implementation, nonadiabatic holonomic quantum computation has received increasing attention.

The seminal scheme of nonadiabatic holonomic quantum computation is based on a three-level quantum system driven by two resonant lasers \cite{Sjoqvist2012,Xu}, where a general one-qubit gate is realized by two-loop implementations. To simplify the operations, the single-shot scheme \cite{Xu2015,Sjovist2016} and the single-loop scheme \cite{Sjovist2016PRA} of nonadiabatic holonomic quantum computation were proposed. The latter two schemes allow us to realize an arbitrary one-qubit gate by a single-shot implementation, which reduces the exposure time of quantum gates to error sources. To have more choices of evolution paths, a general approach of constructing Hamiltonians for nonadiabatic holonomic quantum computation was put forward \cite{Zhao2020}. By this approach, one can find the Hamiltonian that makes the quantum system evolve along a desired path, and thus nonadiabatic holonomic gates can be realized with shortened evolution paths. Up to now, nonadiabatic holonomic quantum computation has been well developed in both theories \cite{Liang,Zhang2014,Mousolou2014,Xue,Zhao,You,Wu,Chen,Zhang2018,Zhao2019,
Zhao2017,Kang,Long2019,Xu2021,Sjoqvist2019,Xing,Xing2020,Liang2022,Zhao2021,Liu} and experiments \cite{Abdumalikov,Long,Arroyo,Duan2014,Long2017,Zhou,Sekiguchi,Xu2018,Danilin,Ishida,Peng,Egger,Yin,Nagata,Ai,XuOP}.

The merit of nonadiabatic holonomic gates comes from their purely geometric property. A crucial step in realizing nonadiabatic holonomic quantum computation is to remove the dynamical phase from the total phase. In the previous schemes, the dynamical phase was removed by resorting to the parallel transport condition, which implies that the instantaneous dynamical phase is always zero. However, this strict requirement limits the realization of nonadiabatic holonomic quantum computation to a special family of quantum systems. Actually, it is not necessary to keep the instantaneous dynamical phase always zero for removing the dynamical phase \cite{Arxiv}.
In this paper, we put forward a strategy to design nonadiabatic holonomic quantum computation, which is based on a commutation relation rather than the parallel transport condition. Instead of requiring the instantaneous dynamical phase to be always zero, the dynamical part of the total phase is separated from the geometric part and then removed by properly choosing evolution parameters. Compared with the previous ones, the schemes based on this strategy are more flexible as the commutation relation is more relaxed than the parallel transport condition. The quantum systems satisfying the commutation relation, containing those satisfying the parallel transport condition as a subset, are more general than the latter.

\section{Strategy}

Consider an $N$-dimensional quantum system governed by the Hamiltonian $H(t)$, of which the evolution operator is denoted as $U(t)=\mathbf{T}\exp[-i\int^t_0 H(t^\prime)dt^\prime]$ with $\mathbf{T}$ being time ordering. We use $\{\ket{\phi_{k}(t)}\}^{N}_{k=1}$ to represent $N$ orthonormal solutions of the Schr\"{o}dinger equation $i\ket{\dot{\phi}_{k}(t)}=H(t)\ket{\phi_{k}(t)}$. Assume there is an $L$-dimensional subspace $\mathcal{S}(t)=\mathrm{Span}\{\ket{\phi_{k}(t)}\}^{L}_{k=1}$
evolving cyclically with the period $\tau$, i.e., $\mathcal{S}(\tau)=\mathcal{S}(0)$. The computational basis can be then encoded into $\mathcal{S}(0)$ and the final evolution operator $U(\tau)$ acting on $\mathcal{S}(0)$ is a quantum gate. $U(\tau)$ plays a holonomic gate if the dynamical part can be removed from it.

To make this point clear, we introduce a set of auxiliary orthonormal bases  $\{\ket{\nu_{k}(t)}\}^{L}_{k=1}$ in the subspace $\mathcal{S}(t)$,  which satisfy  $\ket{\nu_{k}(\tau)}=\ket{\nu_{k}(0)}=\ket{\phi_{k}(0)}$.
Expanding $\ket{\phi_{k}(t)}$ in terms of the auxiliary bases gives
\begin{align}
\ket{\phi_{k}(t)}=\sum^{L}_{l=1}\ket{\nu_{l}(t)}C_{lk}(t),
\label{T1}
\end{align}
where $C_{lk}(t)$ are time-dependent coefficients.
By substituting Eq. (\ref{T1}) into the Schr\"{o}dinger equation, the coefficient matrix can be calculated as
\begin{align}
C(t)=\mathbf{T}e^{i\int^{t}_{0}[A(t^{\prime})-K(t^{\prime})]dt^{\prime}}
\end{align}
with
\begin{align}\label{Tong4}
A_{lk}(t)=i\langle\nu_{l}(t)\ket{\dot{\nu}_{k}(t)},~~~K_{lk}(t)=\bra{\nu_{l}(t)}H(t)\ket{\nu_{k}(t)}.
\end{align}
After a period of time $\tau$, we have
$\ket{\phi_{k}(\tau)}=\sum^{L}_{l=1}\ket{\nu_{l}(\tau)}C_{lk}(\tau)
=\sum^{L}_{l=1}\ket{\phi_{l}(0)}C_{lk}(\tau)$.
Accordingly, the evolution operator acting on the subspace $\mathcal{S}(0)$ is given by
\begin{align}
U(\tau)=C(\tau)=\mathbf{T}e^{i\int^{\tau}_{0}[A(t)-K(t)]dt},
\end{align}
where $A(t)$ and $K(t)$ lead to the geometric and dynamical parts of the evolution operator, respectively.

If the commutation relation,
\begin{align}\label{condition2}
[A(t),K(t^{\prime})]=0,
\end{align}
is fulfilled for $t\in[0,\tau]$  and $t^{\prime}\in[0,\tau]$, the evolution operator can be then written as the product of two parts:
\begin{align}\label{Tong5}
U(\tau)=\left[\mathbf{T}e^{i\int^{\tau}_{0}A(t)dt}\right]\left[\mathbf{T}e^{-i\int^{\tau}_{0}K(t)dt}\right].
\end{align}
The first part $\mathbf{T}\exp[i\int^{\tau}_{0}A(t)dt]$ corresponds to a non-Abelian geometric phase factor and the second part $\mathbf{T}\exp[-i\int^{\tau}_{0}K(t)dt]$ corresponds to a non-Abelian dynamical phase factor. As the dynamical phase factor is separated from the geometric phase factor, it can be removed by letting $\mathbf{T}\exp[-i\int^{\tau}_{0}K(t)dt]=I$. In this case, we have the evolution operator,
\begin{align}\label{Tong7}
U(\tau)=\mathbf{T}e^{i\int^{\tau}_{0}A(t)dt},
\end{align}
which plays a holonomic gate acting on the subspace $\mathcal{S}(0)$.

The key to designing nonadiabatic holonomic quantum computation based on this strategy is to find a quantum system that possesses a cyclically evolutional subspace and satisfies the commutation relation. For this, one can start from the auxiliary bases $\{\ket{\nu_{k}(t)}\}^{L}_{k=1}$. Without loss of generality, we take $N=L+1$ and introduce the $(L+1)$th auxiliary basis $\ket{\nu_{L+1}(t)}=\exp[-i\gamma(t)]\ket{\phi_{L+1}(t)}$, where $\gamma(t)$ is a time-dependent undetermined parameter with $\gamma(0) = 0$.
Since $\ket{\phi_{k}(t)}$ are the solutions of the Schr\"{o}dinger equation
$i\ket{\dot{\phi}_{k}(t)}=H(t)\ket{\phi_{k}(t)}$, the Hamiltonian can be expressed as
\begin{align}
H(t)=i\sum^{L+1}_{k=1}\ket{\dot{\phi}_{k}(t)}\bra{\phi_{k}(t)}.
\label{T2}
\end{align}
Substituting  $\ket{\phi_{k}(t)}=\sum^{L}_{l=1}\ket{\nu_{l}(t)}C_{lk}(t)$ and $\ket{\phi_{L+1}(t)}=\exp[i\gamma(t)]\ket{\nu_{L+1}(t)}$ into Eq. (\ref{T2}), we can obtain
\begin{align}\label{eq1}
H(t)=&i\sum^{L}_{k=1}\langle{\nu_{k}(t)}\ket{\dot{\nu}_{L+1}(t)}\ket{\nu_{k}(t)}\bra{\nu_{L+1}(t)}+\mathrm{H.c.}
\notag\\
&+\sum^{L}_{k,l=1}\bra{\nu_{k}(t)}H(t)\ket{\nu_{l}(t)}\ket{\nu_{k}(t)}\bra{\nu_{l}(t)}
\notag\\
&+[i\langle{\nu_{L+1}(t)}\ket{\dot{\nu}_{L+1}(t)}-\dot{\gamma}(t)]\ket{\nu_{L+1}(t)}\bra{\nu_{L+1}(t)},
\end{align}
where $\mathrm{H.c.}$ represents the Hermitian conjugate terms. Equation (\ref{eq1}) expresses the relation between the Hamiltonian $H(t)$ and the auxiliary bases $\{\ket{\nu_{k}(t)}\}^{L+1}_{k=1}$.
This relation is useful to construct the Hamiltonian for realizing nonadiabatic holonomic quantum gates.

In passing, we would like to point out that the commutation relation $[A(t),K(t^{\prime})]=0$ is naturally satisfied when $K(t)=0$ is taken. In this case, the Hamiltonian in Eq. (\ref{eq1}) is reduced to the special form given in Ref. \cite{Zhao2020}:
\begin{align}\label{Tong3}
H(t)=&i\sum^{L}_{k=1}\langle{\nu_{k}(t)}\ket{\dot{\nu}_{L+1}(t)}\ket{\nu_{k}(t)}\bra{\nu_{L+1}(t)}+\mathrm{H.c.}
\notag\\
&+[i\langle{\nu_{L+1}(t)}\ket{\dot{\nu}_{L+1}(t)}-\dot{\gamma}(t)]\ket{\nu_{L+1}(t)}\bra{\nu_{L+1}(t)}.
\end{align}
Since $K(t)=0$ means that the parallel transport condition, $\bra{\nu_{l}(t)}H(t)\ket{\nu_{k}(t)}=0$ for $l,k=1,\cdots,L$, is fulfilled, we can conclude that the commutation relation is more relaxed than the parallel transport condition. The quantum systems satisfying the commutation relation, containing those satisfying the parallel transport condition as a subset, are more general than the latter, therefore the schemes of nonadiabatic holonomic quantum computation based on the commutation relation are more flexible than those based on the parallel transport condition.

\section{Scheme}

We now show the practicability of our strategy, which is effective to design nonadiabatic holonomic gates indeed.
For a one-qubit nonadiabatic holonomic gate, the quantum system has at least three dimensions, where a two-dimensional subspace is used as a computational space while the remanent one-dimensional subspace acts as an auxiliary space. To this end, we consider a three-level quantum system consisting of two ground states $\ket{0}$ and $\ket{1}$ and an excited state $\ket{e}$.

To construct the quantum system that possesses a cyclically evolutional subspace and satisfies the commutation relation, we take the auxiliary bases as
\begin{align}\label{eq2}
\ket{\nu_{1}(t)}=&\cos\frac{\theta}{2}\ket{0}+\sin\frac{\theta}{2}e^{i\varphi}\ket{1},
\notag\\
\ket{\nu_{2}(t)}=&\cos\frac{\alpha(t)}{2}\sin\frac{\theta}{2}e^{-i\varphi}\ket{0}-\cos\frac{\alpha(t)}{2}\cos\frac{\theta}{2}\ket{1}
\notag\\
\ket{\nu_{3}(t)}=&\sin\frac{\alpha(t)}{2}\sin\frac{\theta}{2}e^{-i[\varphi+\beta(t)]}\ket{0}
-\sin\frac{\alpha(t)}{2}\cos\frac{\theta}{2}e^{-i\beta(t)}\ket{1}
\notag\\
&-\cos\frac{\alpha(t)}{2}\ket{e},
\end{align}
where $\theta$ and $\varphi$ are time-independent parameters, and $\alpha(t)$ and $\beta(t)$ are evolution parameters, being functions of time $t$ with $\alpha(0)=\alpha(\tau)=0$.
One can easily verify that $\mathcal{S}(t)=\mathrm{Span}\{\ket{\nu_{1}(t)},\ket{\nu_{2}(t)}\}$
undergoes cyclic variation such that $\mathcal{S}(\tau)=\mathcal{S}(0)=\mathrm{Span}\{\ket{0},\ket{1}\}$.  Thus, we can take $\{\ket{0},\ket{1}\}$ as the computational basis.

For our purpose, we expect the Hamiltonian to have the form of
\begin{align}\label{eq3}
H(t)=\Delta(t)\ket{e}\bra{e}+[\Omega(t)e^{i\kappa(t)}\ket{e}\bra{b}+\mathrm{H.c.}],
\end{align}
where $\Delta(t)$ is the detuning of lasers, $\Omega(t)$ is the pulse envelope, $\kappa(t)$ is a time-dependent phase, and $\ket{b}=\sin(\theta/2)\exp(-i\varphi)\ket{0}-\cos(\theta/2)\ket{1}$.

To match the Hamiltonian with the auxiliary bases, we substitute Eqs. (\ref{eq2}) and (\ref{eq3}) into Eq. (\ref{eq1}), and compare the coefficients of each term $\ket{i} \bra{j}$ on both sides of the resulting equation \cite{Tonga}. We can obtain
\begin{align}\label{eq6}
&\Delta(t)=-\dot{\alpha}(t)\cot\alpha(t)\cot[\kappa(t)-\beta(t)]-\dot{\beta}(t),
\notag\\
&\Omega(t)e^{i\kappa(t)}=\frac{1}{2}\{i\dot{\alpha}(t)+\dot{\alpha}(t)\cot[\kappa(t)-\beta(t)]\}e^{i\beta(t)},
\end{align}
and $\dot{\gamma}(t)=\dot{\beta}(t)+\dot{\alpha}\cot[\alpha(t)/2]\cot[\kappa(t)-\beta(t)]/2$.

By substituting  Eqs. (\ref{eq2}) and (\ref{eq3}) with the parameters given in Eq. (\ref{eq6}) into Eq. (\ref{Tong4}), a direct calculation shows
\begin{align}\label{Tong8}
A_{11}(t)&=A_{12}(t)=A_{21}(t)=0,
\notag\\
A_{22}(t)&=-\frac{\dot{\beta}(t)}{2}[1-\cos\alpha(t)],
\end{align}
and
\begin{align}\label{eq4}
K_{11}(t)&=K_{12}(t)=K_{21}(t)=0,
\notag\\
K_{22}(t)&=\frac{\dot{\alpha}(t)}{2}\tan\frac{\alpha(t)}{2}\cot[\kappa(t)-\beta(t)]-\dot{\beta}(t)\sin^{2}\frac{\alpha(t)}{2},
\end{align}
One can readily verify that $[A(t),K(t^{\prime})]=0$, i.e., the commutation relation (\ref{condition2}) is fulfilled. It implies that the dynamical phase factor $\mathbf{T}\exp[-i\int^{\tau}_{0}K(t)dt]$ can be extracted from the evolution operator, as shown in Eq. (\ref{Tong5}).

The above discussion shows that the quantum system governed by the Hamiltonian in Eq. (\ref{eq3}) with the parameters given in Eq. (\ref{eq6}) satisfies the requirements that it possesses a cyclically evolutional subspace and satisfies the commutation relation. Such quantum system  is qualified to realize nonadiabatic holonomic quantum computation. We now only need to remove the dynamical part of $U(\tau )$ by letting $\mathbf{T}\exp[-i\int^{\tau}_{0}K(t)dt]=I$. From Eq. (\ref{eq4}), we see that this is guaranteed if
\begin{align}\label{Tong6}
\int^{\tau}_{0}K_{22}(t)dt=0.
\end{align}

Obviously, there are many candidates of $K_{22}(t)$ satisfying Eq. (\ref{Tong6}). For instance, we can choose  $K_{22}(t)=0$.
Alternatively, we can also choose $K_{22}(t)=-\dot{\beta}(t)$ with $\beta(0)=\beta(\tau)$.
In any case, as long as $\int^{\tau}_{0}K_{22}(t)dt=0$, there is $U(\tau)=\mathbf{T}\exp[i\int^{\tau}_{0}A(t)dt]$. Substituting $A(t)$ given in Eq. (\ref{Tong8}) into the integral, we have
\begin{align}\label{eq10}
U(\tau)=\ket{\nu_{1}(0)}\bra{\nu_{1}(0)}+e^{-i\phi(\tau)}\ket{\nu_{2}(0)}\bra{\nu_{2}(0)}
\end{align}
with $\phi(\tau)=\int^{\tau}_{0}\dot{\beta}(t)[1-\cos\alpha(t)]dt/2$.

Ignoring an unimportant global phase, it can be equivalently rewritten as
\begin{align}\label{Tong9}
U(\tau)=e^{i\phi(\tau)\boldsymbol{\mathrm{n}\cdot\sigma}/2},
\end{align}
where $\boldsymbol{\mathrm{n}}=(\sin\theta\cos\varphi,\sin\theta\sin\varphi,\cos\theta)$
and $\boldsymbol{\sigma}=(\sigma_{x},\sigma_{y},\sigma_{z})$. It is an arbitrary one-qubit gate as the direction of the rotation axis $\boldsymbol{\mathrm{n}}$ and the value of the rotation angle $\phi(\tau)$ can be freely chosen. If $\alpha(t)$ and $\beta(t)$ are taken as the polar angle and azimuthal angle of a spherical coordinate system,  $(\alpha(t),\beta(t))$ represents a point in a unit two-sphere. It traces a closed path $C$ in the parameter space when $\alpha(t)$ varies from $\alpha(0)=0$ to $\alpha(\tau)=0$, and $\phi(\tau)$ is just equal to half of the solid angle enclosed by the path $C$:
\begin{align}
\phi(\tau)=\frac{1}{2}\oint_{C}(1-\cos\alpha)d\beta.
\end{align}
Clearly, $\phi(\tau)$ is only dependent on the path in the parameter space but independent of the changing rate of the parameters.

\section{Discussion}

After showing that it is practicable to design nonadiabatic holonomic quantum computation based on the commutation relation, we now discuss some details related to the choice of $K_{22}(t)$ in the above scheme and illustrate the flexibility of our strategy, which allows us to optimize the evolution time and evolution paths.

As stated in the last section, there are many candidates of $K_{22}(t)$ to realize the holonomic gate $U(\tau)$. For instance, it can be taken as $K_{22}(t)=0$ or $K_{22}(t)=-\dot{\beta}(t)$ with $\beta(0)=\beta(\tau)$.
From the expression of $K_{22}(t)$ in Eq. (\ref{eq4}), we see that $K_{22}(t)=0$ means $\dot{\alpha}(t)=\dot{\beta}(t)\sin\alpha(t)\tan[\kappa(t)-\beta(t)]$. Inserting this expression into Eq. (\ref{eq6}), we have $\Delta(t)=-\dot{\beta}(t)[1+\cos\alpha(t)]$ and $\Omega(t)\exp[i\kappa(t)]=[i\dot{\alpha}(t)+\dot{\beta}(t)\sin\alpha(t)]\exp[i\beta(t)]/2$.
Then, the Hamiltonian in Eq. (\ref{eq3}) can be explicitly written as
\begin{align}\label{eq8}
H(t)=&-\dot{\beta}(t)[1+\cos\alpha(t)]\ket{e}\bra{e}
\notag\\
&+\left\{\frac{1}{2}[i\dot{\alpha}(t)+\dot{\beta}(t)\sin\alpha(t)]e^{i\beta(t)}\ket{e}\bra{b}+\mathrm{H.c.}\right\}.
\end{align}
It is just the one given in Ref. \cite{Zhao2020}, some specific expressions of which has been widely used in the previous schemes including the two-loop scheme \cite{Sjoqvist2012,Xu} and one-loop scheme \cite{Sjovist2016PRA}.
Alternatively, if we take $K_{22}(t)=-\dot{\beta}(t)$ with $\beta(0)=\beta(\tau)$, Eq. (\ref{eq4}) results in
$\dot{\alpha}(t)=-2\dot{\beta}(t)\cot[\alpha(t)/2]\cos^{2}[\alpha(t)/2]\tan[\kappa(t)-\beta(t)]$.
Inserting this expression into Eq. (\ref{eq6}), we have
$\Delta(t)=\dot{\beta}(t)\cos\alpha(t)\cot^{2}[\alpha(t)/2]-\dot{\beta}(t)$ and
$\Omega(t)\exp[i\kappa(t)]=\{i\dot{\alpha}(t)-2\dot{\beta}(t)\cot[\alpha(t)/2]\cos^{2}[\alpha(t)/2]\}\exp[i\beta(t)]/2$.
Then, the Hamiltonian in Eq. (\ref{eq3}) can be explicitly written as
\begin{align}\label{eq9}
H(t)=&\left[\dot{\beta}(t)\cos\alpha(t)\cot^{2}\frac{\alpha(t)}{2}-\dot{\beta}(t)\right]\ket{e}\bra{e}+\Bigg\{\frac{1}{2}
\Bigg[i\dot{\alpha}(t)
\notag\\
&-2\dot{\beta}(t)\cot\frac{\alpha(t)}{2}\cos^{2}\frac{\alpha(t)}{2}\Bigg]e^{i\beta(t)}\ket{e}\bra{b}+\mathrm{H.c.}\Bigg\}.
\end{align}
Such Hamiltonian has never been used in previous schemes of nonadiabatic holonomic quantum computation.

\begin{figure}[t]
   \includegraphics[scale=0.34]{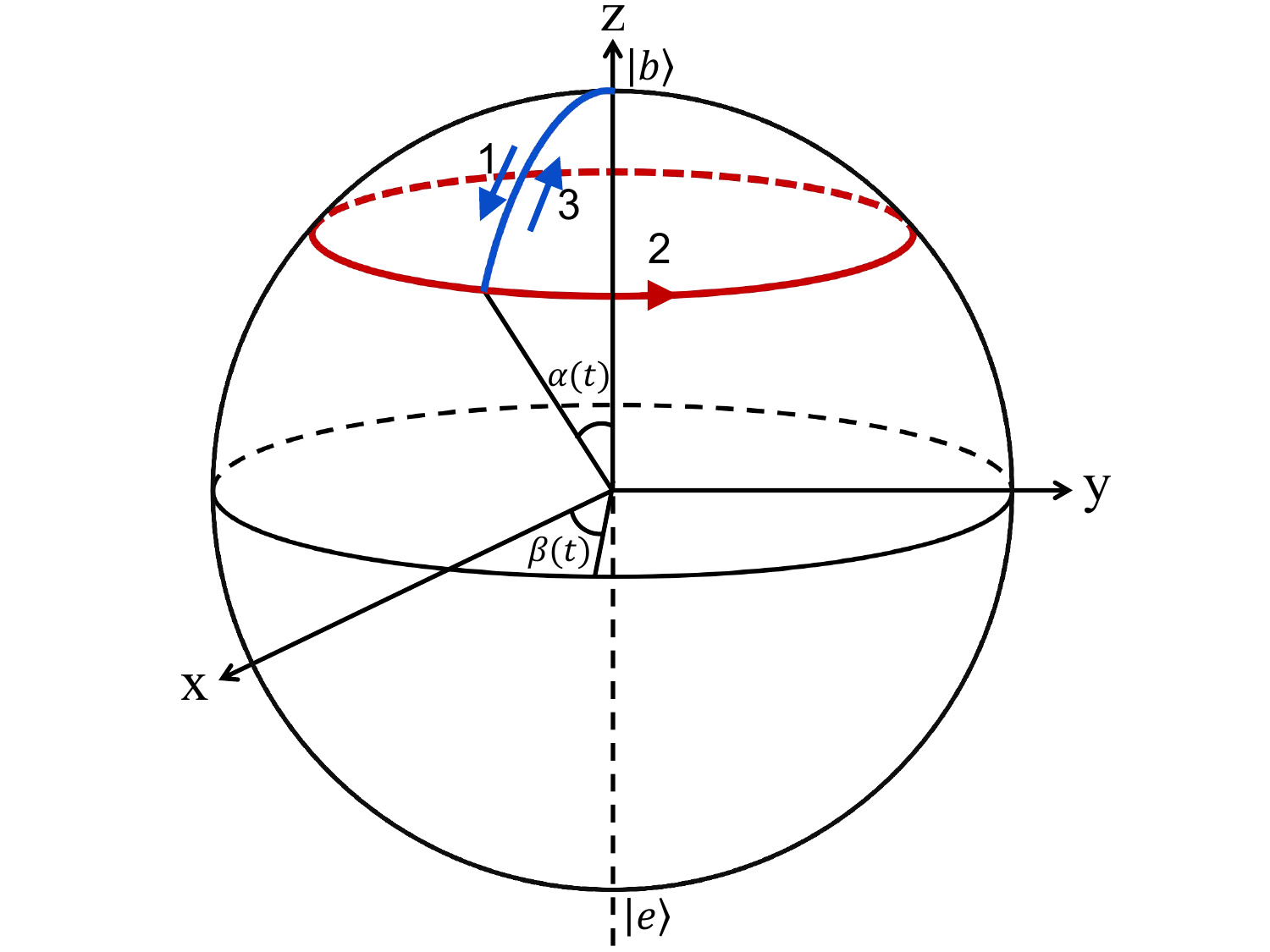}
   \caption{The Bloch sphere representation of the evolution path that starts from the north pole along the great circle with $\beta(t)=0$ to the point $(\alpha,0)$, then along the circle with $\alpha(t)=\alpha$ for a round, and finally along the great circle with $\beta(t)=0$ to the north pole.}
   \label{Fig1}
\end{figure}

Each choice of $K_{22}(t)$ has its own advantages and we can optimize the realization by a proper choice. To illustrate this point, we compare the evolution time in the above two choices by using  the pulse area as a measure of evolution time. We take the evolution path which starts from the north pole along the great circle with $\beta(t)=0$ to the point $(\alpha,0)$, then along the circle with $\alpha(t)=\alpha$ for a round, and finally along the great circle with $\beta(t)=0$ to the north pole. That is, the path consists of three segments, as shown in Fig. \ref{Fig1}, corresponding to the time intervals $t\in[0,\tau_{1}]$, $t\in(\tau_{1},\tau_{2}]$ and $t\in(\tau_{2},\tau]$, respectively.

If we use the Hamiltonian in Eq. (\ref{eq8}), which corresponds to $K_{22}=0$, to realize the nonadiabatic holonomic gate,
its piecewise expression reads $H(t)=i\dot{\alpha}(t)/2\ket{e}\bra{b}+\mathrm{H.c.}$ for $t\in[0,\tau_{1}]$, ~$-\dot{\beta}(t)(1+\cos\alpha)\ket{e}\bra{e}+[\dot{\beta}(t)\sin\alpha\exp[i\beta(t)]/2\ket{e}\bra{b}+\mathrm{H.c.}]$ for  $t\in(\tau_{1},\tau_{2}]$, and $i\dot{\alpha}(t)/2\ket{e}\bra{b}+\mathrm{H.c.}$ for  $t\in(\tau_{2},\tau]$.  Therefore, the pulse envelope, denoted as $\Omega(t)$, is  $\Omega(t)=\dot{\alpha}(t)/2$ for $t\in[0,\tau_{1}]$,~ $[\dot{\beta}(t)\sin\alpha]/2$ for  $t\in(\tau_{1},\tau_{2}]$, and $-\dot{\alpha}(t)/2$  for  $t\in(\tau_{2},\tau]$. We then can calculate the pulse area as $\mathcal{A}_1=\int^{\tau}_{0}\Omega(t)dt=\pi\sin\alpha+\alpha$.

If we use the Hamiltonian in Eq. (\ref{eq9}), which corresponds to $K_{22}(t)=-\dot{\beta}(t)$, to realize the nonadiabatic holonomic gate, its piecewise expression is $H(t)=i\dot{\alpha}(t)/2\ket{e}\bra{b}+\mathrm{H.c.}$ for  $t\in[0,\tau_1]$, ~ $\dot{\beta}(t)[\cos\alpha\cot^{2}(\alpha/2)-1]\ket{e}\bra{e}-\{\dot{\beta}(t)\cot(\alpha/2)
\cos^{2}(\alpha/2)\exp[i\beta(t)]\ket{e}\bra{b}+\mathrm{H.c.}\}$ for $t\in(\tau_1,\tau_2]$, and $i\dot{\alpha}(t)/2\ket{e}\bra{b}+\mathrm{H.c.}$ for $t\in(\tau_{2},\tau]$. Correspondingly, the pulse envelope reads $\Omega(t)= \dot{\alpha}(t)/2$ for  $t\in[0,\tau_1]$, ~ $\dot{\beta}(t)\cot(\alpha/2)\cos^{2}(\alpha/2)$  for  $t\in(\tau_1,\tau_2]$, and  $-\dot{\alpha}(t)/2$  for  $t\in(\tau_2,\tau]$.
In this case, the pulse area reads $\mathcal{A}_2=\int^{\tau}_{0}\Omega(t)dt=2\pi\cot(\alpha/2)\cos^{2}(\alpha/2)+\alpha$.

Comparing the two cases, we have $\mathcal{A}_1<\mathcal{A}_2$ for $\alpha\in[0,\pi/2)$ and $\mathcal{A}_2<\mathcal{A}_1$ for $\alpha\in(\pi/2,\pi]$. Since $\phi(\tau)=\pi(1-\cos\alpha)$, we see that it needs a shorter time to realize the quantum gate with $\phi(\tau)\in[0,\pi)$ by using the Hamiltonian in Eq. (\ref{eq8}) than that in Eq. (\ref{eq9}). Conversely, it needs a shorter time to realize the quantum gate with $\phi(\tau)\in(\pi,2\pi]$ by using the Hamiltonian in Eq. (\ref{eq9}) than that in Eq. (\ref{eq8}). Therefore, our strategy allows us to optimize the evolution time of realizing a nonadiabatic holonomic gate.

Similarly, we can demonstrate that our strategy also allows us to optimize the evolution path of realizing a nonadiabatic holonomic gate, as many paths can be chosen for realizing the same nonadiabatic holonomic gate. For example, when using resonant coupling to realize nonadiabatic holonomic gates, the previous schemes based on the parallel transport condition only permit us to take the great circle and orange-slice-shaped loop, while the scheme based on the commutation relation permits us to take many available paths but not limited to the great circle and orange-slice-shaped loop. Here, resonant coupling means that the detuning of lasers is equal to zero.

The Hamiltonian in Eqs. (\ref{eq8}) and (\ref{eq9}) is achievable in physical systems. To see this, we can generally express them as the form $H(t)=\Delta(t)\ket{e}\bra{e}+[\tilde{\Omega}(t)\ket{e}\bra{b}+\mathrm{H.c.}]$. By substituting $\ket{b}=\sin(\theta/2)\exp(-i\varphi)\ket{0}-\cos(\theta/2)\ket{1}$ into the expression, $H(t)$ can be further written as $H(t)=\Delta(t)\ket{e}\bra{e}+[\tilde{\Omega}(t)\sin(\theta/2)\exp(i\varphi)\ket{e}\bra{0}
-\tilde{\Omega}(t)\cos(\theta/2)\ket{e}\bra{1}+\mathrm{H.c.}]$.
Obviously, such Hamiltonian describes a three-level quantum system driven by two off-resonant lasers with common detuning $\Delta(t)$ and different Rabi frequencies $\tilde{\Omega}(t)\sin(\theta/2)\exp(i\varphi)$ and $-\tilde{\Omega}(t)\cos(\theta/2)$. It can be implemented in many physical systems, such as superconducting circuits \cite{Yin} and nitrogen–vacancy centers in diamond \cite{Duan2019}. Here, $\theta$ and $\varphi$ completely determine the direction of the rotation axis $\boldsymbol{\mathrm{n}}$, and they are constants for a specific quantum gate. $\Delta(t)$ and $\tilde{\Omega}(t)$ are determined by $\alpha(t)$ and $\beta(t)$.
For a given evolution path traced by $(\alpha(t),\beta(t))$, $\Delta(t)$ and $\tilde{\Omega}(t)$ can be fixed and thus $H(t)$ is completely determined. For example, if the evolution path traced by $(\alpha(t),\beta(t))$ is taken as the one in Fig. \ref{Fig1}, the Hamiltonian in Eq. (\ref{eq8}) yields $H(t)=i\dot{\alpha}(t)/2\ket{e}\bra{b}+\mathrm{H.c.}$ for $t\in[0,\tau_{1}]$, ~$-\dot{\beta}(t)(1+\cos\alpha)\ket{e}\bra{e}+[\dot{\beta}(t)\sin\alpha\exp[i\beta(t)]/2\ket{e}\bra{b}+\mathrm{H.c.}]$ for  $t\in(\tau_{1},\tau_{2}]$, and $i\dot{\alpha}(t)/2\ket{e}\bra{b}+\mathrm{H.c.}$ for  $t\in(\tau_{2},\tau]$.
Here, $\Delta(t)=0$ and $\tilde{\Omega}(t)=i\dot{\alpha}(t)/2$ for $t\in[0,\tau_{1}]\cup(\tau_{2},\tau]$, and $\Delta(t)=-\dot{\beta}(t)(1+\cos\alpha)$ and $\tilde{\Omega}(t)=\dot{\beta}(t)\sin\alpha\exp[i\beta(t)]/2$ for $t\in(\tau_{1},\tau_{2}]$.

\begin{figure}[t]
   \includegraphics[scale=0.58]{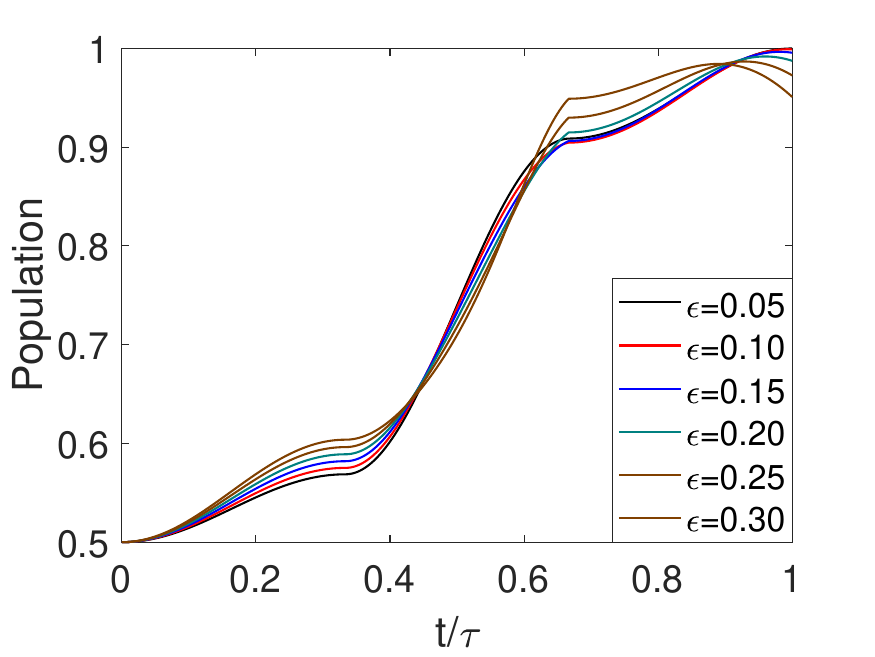}
   \caption{The population $P(t)=|\langle{\phi_{d}}\ket{\phi_{r}(t)}|^{2}$ as a function of $t/\tau$, where $\ket{\phi_{r}(t)}$ is the real evolution state at time $t$. The fidelity is equal to the population at the final time $\tau$, i.e., $F=P(\tau)$.}
   \label{Fig2}
\end{figure}

The success of our scheme is dependent on the condition stated in Eq. (\ref{Tong6}), which guarantees the removal of the dynamical part. If the Hamiltonian of the quantum system is exactly controlled, the condition determined by $(\alpha(t),\beta(t))$ is strictly satisfied and the holonomic gate can be accurately realized. However, if the control parameters, such as $\Delta(t)$ and $\tilde{\Omega}(t)$ in the Hamiltonian, contain errors due to some inevitable noises, the evolution path traced by $(\alpha(t),\beta(t))$ will not be the desired one and thus the accuracy of the quantum gate may be affected. For this, we numerically simulate the performance of our scheme under imperfect parameters $\Delta(t)$ and $\tilde{\Omega}(t)$ by taking the widely used Hadamard gate $H$ as an example. The Hadamard gate $H$ is a rotation operation with the rotation axis $(\sigma_{x}+\sigma_{z})/\sqrt{2}$ and rotation angle $\pi$, which correspond to $\theta=\pi/4$, $\varphi=0$, and $\phi(\tau)=\pi$.
For our purpose, we take the input state as $\ket{0}$, the Hamiltonian as that in Eq. (\ref{eq8}), and the
evolution path traced by $(\alpha(t),\beta(t))$ as the one in Fig. \ref{Fig1}.
Furthermore, we set
$\alpha(t)=\pi\sin[\pi{t}/(2\tau_{1})]/2$ and $\beta(t)=0$ for $t\in[0,\tau_{1}]$,
$\alpha(t)=\pi/2$ and $\beta(t)=2\pi(t-\tau_{1})/(\tau_{2}-\tau_{1})$ for $t\in(\tau_{1},\tau_{2}]$, and
$\alpha(t)=\pi\sin\{\pi(\tau-t)/[2(\tau-\tau_{2})]\}/2$ and $\beta(t)=0$ for $t\in(\tau_{2},\tau]$.
Then, we have $\phi(\tau)=\oint_{C}(1-\cos\alpha)d\beta/2=\pi$ and thus the Hadamard gate $H$ can be realized.
In this case, the parameters should be taken as $\Delta(t)=0$ and $\tilde{\Omega}(t)=i\pi^{2}\cos[\pi{t}/(2\tau_{1})]/(8\tau_{1})$ for  $t\in[0,\tau_{1}]$, $\Delta(t)=-2\pi/(\tau_{2}-\tau_{1})$ and $\tilde{\Omega}(t)=\pi\exp[i2\pi(t-\tau_{1})/(\tau_{2}-\tau_{1})]/(\tau_{2}-\tau_{1})$ for  $t\in[\tau_{1},\tau_{2}]$, and $\Delta(t)=0$ and $\tilde{\Omega}(t)=-i\pi^{2}\cos\{\pi(\tau-t)/[2(\tau-\tau_{2})\}/[8(\tau-\tau_{2})]$ for  $t\in[\tau_{2},\tau]$.
Let us now assume that there exist systematic errors for the parameters such that $\Delta(t)\rightarrow(1+\epsilon)\Delta(t)$ and $\tilde{\Omega}(t)\rightarrow(1+\epsilon)\tilde{\Omega}(t)$, where $\epsilon$ is a small number.
With the aid of numerical simulation, we calculate the fidelity $F=|\langle{\phi_{d}}\ket{\phi_{r}}|^{2}$ between the desired output state $\ket{\phi_{d}}$ and the real output state $\ket{\phi_{r}}$. The result indicates that the fidelities corresponding to $\epsilon=0.05$, $0.10$, $0.15$, $0.20$, $0.25$, and $0.30$ can be up to $99.99\%$, $99.91\%$, $99.56\%$, $98.74\%$ ,$97.27\%$, and $95.09\%$, respectively, as depicted in Fig. \ref{Fig2}.

Besides arbitrary one-qubit gates, a nontrivial two-qubit gate is also needed for nonadiabatic holonomic quantum computation. To realize a two-qubit gate, one can take the auxiliary bases as
\begin{align}
\ket{\nu_{1}(t)}=&\ket{00},~~~\ket{\nu_{2}(t)}=\ket{01},
\notag\\
\ket{\nu_{3}(t)}=&\cos\frac{\theta}{2}\ket{10}+\sin\frac{\theta}{2}e^{i\varphi}\ket{11},
\notag\\
\ket{\nu_{4}(t)}=&\cos\frac{\alpha(t)}{2}\sin\frac{\theta}{2}e^{-i\varphi}\ket{10}
-\cos\frac{\alpha(t)}{2}\cos\frac{\theta}{2}\ket{11},
\notag\\
\ket{\nu_{4}(t)}=&\cos\frac{\alpha(t)}{2}\sin\frac{\theta}{2}e^{-i\varphi}\ket{10}
-\cos\frac{\alpha(t)}{2}\cos\frac{\theta}{2}\ket{11}
\notag\\
&+\sin\frac{\alpha(t)}{2}e^{i\beta(t)}\ket{ee},
\notag\\
\ket{\nu_{5}(t)}=&\sin\frac{\alpha(t)}{2}\sin\frac{\theta}{2}e^{-i[\varphi+\beta(t)]}\ket{10}
-\sin\frac{\alpha(t)}{2}\cos\frac{\theta}{2}e^{-i\beta(t)}\ket{11}
\notag\\
&-\cos\frac{\alpha(t)}{2}\ket{ee},
\end{align}
where $\theta$ and $\varphi$ are time-independent parameters and $\alpha(t)$ and $\beta(t)$ are time-dependent parameters with $\alpha(0)=\alpha(\tau)=0$. Note that $\{\ket{\nu_{1}(t)},\ket{\nu_{2}(t)}\}$ are invariant and
$\{\ket{\nu_{3}(t)},\ket{\nu_{4}(t)},\ket{\nu_{5}(t)}\}$ have the same form as Eq. (\ref{eq2}).
Therefore, we can use a similar approach to one-qubit gates to realize the two-qubit gate.

So far, we have demonstrated that besides requiring the instantaneous dynamical part to be always zero, nonadiabatic holonomic quantum computation can be also realized by separating the dynamical part of the total phase from the geometric part and then removing the dynamical part. This is similar to the case of nonadiabatic geometric quantum computation \cite{Wang,Zhu2002}, where the dynamical phase can be removed by requiring the instantaneous dynamical part to be always zero or by a dynamical compensation method with multiple evolution paths \cite{Zhu2003}. Furthermore, nonadiabatic geometric quantum computation can be realized too without removing the dynamical phase but with requiring the dynamical phase to be proportional to the geometric phase \cite{ZhuPRL,Du}. However, it is an open problem for realizing nonadiabatic holonomic quantum computation without removing the dynamical phase.

\section{Conclusion}

In conclusion, by introducing the commutation relation defined in Eq. (\ref{condition2}), we put forward a strategy to design nonadiabatic holonomic quantum computation. The key to realizing a nonadiabatic holonomic gate based on this strategy is to construct the Hamiltonian of the quantum system that possesses a cyclically evolutional subspace and satisfies the commutation relation. The commutation relation guarantees that the dynamical part of the evolution operator is separated from the geometric part, which can be removed by properly choosing evolution parameters. To show the practicability of our strategy, a set of Hamiltonians that can realize nonadiabatic holonomic quantum computation is given too.

The schemes of nonadiabatic holonomic quantum computation based on the commutation relation are more flexible than the previous ones as the commutation relation is more relaxed than the parallel transport condition. The quantum systems satisfying the commutation relation, containing those satisfying the parallel transport condition as a subset, are more general than the latter. They provide more options for realizing nonadiabatic holonomic quantum computation, and hence allow us to optimize realizations such as the evolution time and evolution paths.

\begin{acknowledgments}
We acknowledge support from the National Natural Science Foundation of China through Grant No. 12174224.
\end{acknowledgments}


\begin{thebibliography}{99}

\bibitem{Berry} M. V. Berry, Proc. R. Soc. Lond. A \textbf{392}, 45 (1984).
\bibitem{Wilczek} F. Wilczek and A. Zee, Phys. Rev. Lett. \textbf{52}, 2111 (1984).
\bibitem{Aharonov} Y. Aharonov and J. Anandan, Phys. Rev. Lett. \textbf{58}, 1593 (1987).
\bibitem{Anandan} J. Anandan, Phys. Lett. A \textbf{133}, 171 (1988).
\bibitem{Jones} J. A. Jones, V. Vedral, A. Ekert, and G. Castagnoli, Nature \textbf{403}, 869 (2000).
\bibitem{Zanardi} P. Zanardi and M. Rasetti, Phys. Lett. A \textbf{264}, 94 (1999).
\bibitem{Duan} L. M. Duan, J. I. Cirac, and P. Zoller, Science \textbf{292}, 1695 (2001).
\bibitem{Wang} Xiang-Bin Wang and Matsumoto Keiji, Phys. Rev. Lett. \textbf{87}, 097901 (2001).
\bibitem{Zhu2002} S. L. Zhu and Z. D. Wang, Phys. Rev. Lett. \textbf{89}, 097902 (2002).
\bibitem{Sjoqvist2012} E. Sj\"{o}qvist, D. M. Tong, L. M. Andersson, B. Hessmo, M. Johansson, and K. Singh, New J. Phys. \textbf{14}, 103035 (2012).
\bibitem{Xu} G. F. Xu, J. Zhang, D. M. Tong, E. Sj\"{o}qvist, and L. C. Kwek, Phys. Rev. Lett. \textbf{109}, 170501 (2012).
\bibitem{Xu2015} G. F. Xu, C. L. Liu, P. Z. Zhao, and D. M. Tong, Phys. Rev. A \textbf{92}, 052302 (2015).
\bibitem{Sjovist2016} E. Sj\"{o}qvist, Phys. Lett. A \textbf{380}, 65 (2016).
\bibitem{Sjovist2016PRA} E. Herterich and E. Sj\"{o}qvist, Phys. Rev. A \textbf{94}, 052310 (2016).
\bibitem{Zhao2020} P. Z. Zhao, K. Z. Li, G. F. Xu, and D. M. Tong, Phys. Rev. A \textbf{101}, 062306 (2020).
\bibitem{Liang} Z. T. Liang, Y. X. Du, W. Huang, Z. Y. Xue, and H. Yan, Phys. Rev. A \textbf{89}, 062312 (2014).
\bibitem{Zhang2014} J. Zhang, L. C. Kwek, E. Sj\"oqvist, D. M. Tong, and P. Zanardi, Phys. Rev. A \textbf{89}, 042302 (2014).
\bibitem{Mousolou2014} V. A. Mousolou, C. M. Canali, and E. Sj\"{o}qvist, New J. Phys. \textbf{16}, 013029 (2014).
\bibitem{Xue} Z. Y. Xue, J. Zhou, and Z. D. Wang, Phys. Rev. A \textbf{92}, 022320 (2015).
\bibitem{You} Y. M. Wang, J. Zhang, C. F. Wu, J. Q. You, and G. Romero, Phys. Rev. A \textbf{94}, 012328 (2016).
\bibitem{Zhao2017} P. Z. Zhao, G. F. Xu, Q. M. Ding, E. Sj\"{o}qvist, and D. M. Tong, Phys. Rev. A \textbf{95}, 062310 (2017).
\bibitem{Zhao} P. Z. Zhao, X. D. Cui, G. F. Xu, E. Sj\"{o}qvist, and D. M. Tong, Phys. Rev. A \textbf{96}, 052316 (2017).
\bibitem{Kang} Y. H. Kang, Y. H. Chen, Z. C. Shi, B. H. Huang, J. Song, and Y. Xia, Phys. Rev. A \textbf{97}, 042336 (2018).
\bibitem{Zhang2018} J. Zhang, S. J. Devitt, J. Q. You, and F. Nori, Phys. Rev. A \textbf{97}, 022335 (2018).
\bibitem{Chen} T. Chen, J. Zhang, and Z. Y. Xue, Phys. Rev. A \textbf{98}, 052314 (2018).
\bibitem{Zhao2019} P. Z. Zhao, G. F. Xu, and D. M. Tong, Phys. Rev. A \textbf{99}, 052309 (2019).
\bibitem{Sjoqvist2019} N. Ramberg and E. Sj\"{o}qvist, Phys. Rev. Lett. \textbf{122}, 140501 (2019).
\bibitem{Liu} B. J. Liu, X. K. Song, Z. Y. Xue, X. Wang, and M. H. Yung, Phys. Rev. Lett. \textbf{123}, 100501 (2019).
\bibitem{Long2019} F. H. Zhang, J. Zhang, P. Gao, and G. L. Long, Phys. Rev. A \textbf{100}, 012329 (2019).
\bibitem{Wu} C. F. Wu, Y. M. Wang, X. L. Feng, and J. L. Chen, Phys. Rev. Applied \textbf{13}, 014055 (2020).
\bibitem{Xing2020} T. H. Xing, X. Wu, and G. F. Xu, Phys. Rev. A \textbf{101}, 012306 (2020).
\bibitem{Xing} T. H. Xing and D. M. Tong, Chin. Sci. Bull. \textbf{65}, 2499 (2020).
\bibitem{Xu2021} G. F. Xu, P. Z. Zhao, Erik Sj\"{o}qvist, and D. M. Tong, Phys. Rev. A \textbf{103}, 052605 (2021).
\bibitem{Zhao2021} P. Z. Zhao, G. F. Xu, and D. M. Tong, Chin. Sci. Bull. \textbf{66}, 1935 (2021).
\bibitem{Liang2022} Y. Liang, P. Shen, T. Chen, and Z. Y. Xue, Phys. Rev. Appl. \textbf{17}, 034015 (2022).
\bibitem{Abdumalikov} A. A. Abdumalikov, J. M. Fink, K. Juliusson, M. Pechal, S. Berger, A. Wallraff, and S. Filipp, Natur \textbf{496}, 482 (2013).
\bibitem{Long} G. R. Feng, G. F. Xu, and G. L. Long, Phys. Rev. Lett. \textbf{110}, 190501 (2013).
\bibitem{Duan2014} C. Zu, W. B. Wang, L. He, W. G. Zhang, C. Y. Dai, F. Wang, and L. M. Duan, Nature (London) \textbf{514}, 72 (2014).
\bibitem{Arroyo} S. A. Camejo, A. Lazariev, S. W. Hell, and G. Balasubramanian, Nat. Commun. \textbf{5}, 4870 (2014).
\bibitem{Zhou} B. B. Zhou, P. C. Jerger, V. O. Shkolnikov, F. J. Heremans, G. Burkard, and D. D. Awschalom, Phys. Rev. Lett. \textbf{119}, 140503 (2017).
\bibitem{Sekiguchi} Y. Sekiguchi, N. Niikura, R. Kuroiwa, H. Kano, and H. Kosaka, Nat. Photonics {\bf 11}, 309 (2017).
\bibitem{Long2017} H. Li, Y. Liu, and G. L. Long, Sci. China: Phys., Mech. Astron. \textbf{60}, 080311 (2017).
\bibitem{Xu2018} Y. Xu, W. Cai, Y. Ma, X. Mu, L. Hu, Tao Chen, H. Wang, Y. P. Song, Z. Y. Xue, Z. Q. Yin, and L. Sun, Phys. Rev. Lett. \textbf{121}, 110501 (2018).
\bibitem{Nagata} K. Nagata, K. Kuramitani, Y. Sekiguchi, and H. Kosaka, Nat. Commun. {\bf 9}, 3227 (2018).
\bibitem{Ishida} N. Ishida, T. Nakamura, T. Tanaka, S. Mishima, H. Kano, R. Kuroiwa, Y. Sekiguchi, and H. Kosaka, Opt. Lett. {\bf 43}, 2380 (2018).
\bibitem{Danilin} S. Danilin, A. Veps\"{a}l\"{a}inen, and G. S. Paraoanu, Phys. Scr. \textbf{93}, 055101 (2018).
\bibitem{Peng} Z. N. Zhu, T. Chen, X. D. Yang, J. Bian, Z. Y. Xue, and X. H. Peng, Phys. Rev. Appl. \textbf{12}, 024024 (2019).
\bibitem{Egger} D. J. Egger, M. Ganzhorn, G. Salis, A. Fuhrer, P. M\"{u}ller, P. K. Barkoutsos, N. Moll, I. Tavernelli, and S. Filipp, Phys. Rev. Applied \textbf{11}, 014017 (2019).
\bibitem{Yin} Z. X. Zhang, P. Z. Zhao, T. H. Wang, L. Xiang, Z. L. Jia, P. Duan, D. M. Tong, Y. Yin and G. P. Guo, New J. Phys. \textbf{21}, 073024 (2019)
\bibitem{Ai} M. Z. Ai, S. Li, Z. B. Hou, R. He, Z. H Qian, Z. Y. Xue, J. M. Cui, Y. F. Huang, C. F. Li, and G. C. Guo. Phys. Rev. Appl. \textbf{14}, 054062 (2020).
\bibitem{XuOP} K. Xu, W. Ning, X. J. Huang, P. R. Han, H. K. Li, Z. B. Yang, D. N. Zheng, H. Fan, and S. B. Zheng, Optica \textbf{8}, 972 (2021).
\bibitem{Arxiv} J. Zhang, T. H. Kyaw, S. Filipp, L. C. Kwek, E. Sj\"{o}qvist, and D. M. Tong, arXiv:2110.03602.
\bibitem{Tonga}
By Substituting Eqs. (\ref{eq2}) and (\ref{eq3}) into the both sides of (\ref{eq1}), the resulting equation is
$\Delta(t)\ket{e}\bra{e}+[\Omega(t)e^{i\kappa(t)}\ket{e}\bra{b}+\mathrm{H.c.}]
=\{i\dot{\alpha}(t)/2-\dot{\beta}(t)\sin\alpha(t)[1+\cos\alpha(t)]/4
+\dot{\gamma}(t)\sin\alpha(t)/2+\Delta(t)\sin\alpha(t)[1-\cos\alpha(t)]/4
+\Omega(t)\sin^{2}\alpha(t)\cos[\kappa(t)-\beta(t)]/2\}\exp[i\beta(t)]
\ket{e}\bra{b} +\mathrm{H.c.}
-\{\dot{\beta}(t)\sin^{2}\alpha(t)/4
+\dot{\gamma}(t)[1+\cos\alpha(t)]/2-\Delta(t)[1-\cos\alpha(t)]^{2}/4
-\Omega(t)\sin\alpha(t)[1-\cos\alpha(t)]\cos[\kappa(t)-\beta(t)]/2\}\ket{e}\bra{e}
+\{\dot{\beta}[1-\cos\alpha(t)][3+\cos\alpha(t)]/4
-\dot{\gamma}(t)[1-\cos\alpha(t)]/2+\Delta(t)\sin^{2}\alpha(t)/4
+\Omega(t)\sin\alpha(t)[1+\cos\alpha(t)]\cos[\kappa(t)-\beta(t)]/2\}\ket{b}\bra{b}$.
\bibitem{Duan2019} Y. Y. Huang, Y. K. Wu, F. Wang, P. Y. Hou, W. B. Wang, W. G. Zhang, W. Q. Lian, Y. Q. Liu, H. Y. Wang, H. Y. Zhang, L. He, X. Y. Chang, Y. Xu, and L. M. Duan, Phys. Rev. Lett. \textbf{122}, 010503 (2019).
\bibitem{Zhu2003} S. L. Zhu and Z. D. Wang, Phys. Rev. A \textbf{67}, 022319 (2003).
\bibitem{ZhuPRL} S. L. Zhu and Z. D. Wang, Phys. Rev. Lett. \textbf{91}, 187902 (2003).
\bibitem{Du} J. F. Du, P. Zou, and Z. D. Wang, Phys. Rev. A \textbf{74}, 020302(R) (2006).

\end{thebibliography}
\end{document}